\documentstyle[fleqn,11pt]{book}
\begin{document}
\parindent 1.4cm
\large
\begin{center}
{THE QUANTUM WAVE PACKET AND THE FEYNMAN DE BROGLIE BOHM
PROPAGATOR OF THE LINEARIZED SUSSMANN HASSE ALBRECHT KOSTIN NASSAR
EQUATION ALONG A CLASSICAL TRAJETORY}
\end{center}
\begin{center}
{J. M. F. Bassalo$^{1}$,\ P. T. S. Alencar$^{2}$,\  D. G. da
Silva$^{3}$,\ A. B. Nassar$^{4}$\ and\ M. Cattani$^{5}$}
\end{center}
\begin{center}
{$^{1}$\ Funda\c{c}\~ao Minerva,\ Avenida Governador Jos\'e
Malcher\ 629 -\ CEP\ 66035-100,\ Bel\'em,\ Par\'a,\ Brasil
E-mail:\ jmfbassalo@gmail.com}
\end{center}
\begin{center}
{$^{2}$\ Universidade Federal do Par\'a\ -\ CEP\ 66075-900,\
Guam\'a, Bel\'em,\ Par\'a,\ Brasil E-mail:\ tarso@ufpa.br}
\end{center}
\begin{center}
{$^{3}$\ Escola Munguba do Jari, Vit\'oria do Jari\ -\ CEP\
68924-000,\ Amap\'a,\ Brasil E-mail:\ danielgemaque@yahoo.com.br}
\end{center}
\begin{center}
{$^{4}$\ Extension Program-Department of Sciences, University of
California,\ Los Angeles, California 90024,\ USA E-mail:\
nassar@ucla.edu}
\end{center}
\begin{center}
{$^{5}$\ Instituto de F\'{\i}sica da Universidade de S\~ao Paulo.
C. P. 66318, CEP\ 05315-970,\ S\~ao Paulo,\ SP, Brasil E-mail:\
mcattani@if.usp.br}
\end{center}
\par
Abstract:\ In this paper we study the quantum wave
packet and the Feynman-de Broglie-Bohm propagator of the
linearized S\"{u}ssmann-Hasse-Albrecht-Kostin-Nassar equation along a
classical trajetory.
\vspace{0.2cm}
\par
PACS\ 03.65\ -\ Quantum Mechanics
\par
\vspace{0.2cm}
\par
1.\ {\bf Introduction}
\vspace{0.2cm}
\par
In the present work we investigate the quantum wave packet and the
Feynman-de Broglie-Bohm propagator of the linearized
S\"{u}ssmann-Hasse-Albrecht-Kostin-Nassar equation along a classical
trajetory by using the Quantum Mechanical of the de
Broglie-Bohm.$^{[1]}$
\vspace{0.2cm}
\par
2.\ {\bf The S\"{u}ssmann-Hasse-Albrecht-Kostin-Nassar Equation}
\par
In 1973,$^{[2]}$ D. S\"{u}ssmann, and in 1975, R. W. Hasse,$^{[3]}$ K.
Albrecht$^{[4]}$ and M. D. Kostin$^{[5]}$ proposed a non-linear
Schr\"{o}dinger Equation, that was generalized by A. B. Nassar, in
1986,$^{[6]}$ to represent time dependent physical systems, given by:
\begin{center}
{$i\ {\hbar}\ {\frac {{\partial}}{{\partial}t}}\ {\psi}(x,\ t)\ =\ -\
{\frac {{\hbar}^{2}}{2\ m}}\ {\frac {{\partial}^{2}{\psi}(x,\
t)}{{\partial}x^{2}}}\ +$}
\end{center}
\begin{center}
{$+\ {\Big {[}}\ V(x,\ t)\ +\ {\nu}\ {\Big {(}}\ [x\ -\ q(t)]\ [c\ {\hat
{p}}\ +\ (1\ -\ c)\ <\ {\hat {p}}\ >]\ -\ {\frac {i\ {\hbar}\ c}{2}} {\Big
{)}}\ {\Big {]}}\ {\psi}(x,\ t)$\ ,\ \ \ \ \ (2.1)}
\end{center}
where ${\hat {p}}$ is the operator of linear momentum:
\begin{center}
{$\ {\hat {p}}\ =\ -\ i\ {\hbar}\ {\frac {{\partial}}{{\partial}\ x}}$\ ,\
\ \ \ \ (2.2)}
\end{center}
and $c$ is a constant, where:\ $c\ =\ 1$, for S\"{u}ssmann;\ $c\ =\
1/2$, for Hasse; and $c\ =\ 0$, for Albrecht and Kostin. Besides,
${\psi}(x,\ t)$ and $V(x,\ t)$ are, respectively, the wavefunction and
the time dependent potential of the physical system in study, $q(t)\ =\
<\ x>$, and ${\nu}$ is a constant.
\par
Writting the wavefunction ${\psi}(x,\ t)$ in the polar form, defined by
the Madelung-Bohm:$^{[7,\ 8]}$
\begin{center}
{${\psi}(x,\ t)\ =\ {\phi}(x,\ t)\ exp\ [i\ S(x,\ t)]$,\ \ \ \ \ (2.3)}
\end{center}
where $S(x,\ t)$ is the classical action and ${\phi}(x,\ t)$ will be
defined in what follows. Now, using the eq. (2.2) into eq. (2.3), we
get:\ $^{[1]}$
\par
\begin{center}
{${\hat {p}}\ {\psi}\ =\ -\ i\ {\hbar}\ {\frac
{{\partial}}{{\partial}x}}\ ({\phi}\ e^{i\ S})\ =\ -\ i\ {\hbar}\
({\frac {1}{{\phi}}}\ {\frac {{\partial}{\phi}}{{\partial}x}}\ +\ i\
{\frac {{\partial}S}{{\partial}x}})\ {\psi}\ \ \ {\to}$}
\end{center}
\begin{center}
{${\hat {p}}\ {\psi}\ =\ {\hbar}\ ({\frac {{\partial}S}{{\partial}x}}\
-\ {\frac {i}{{\phi}}}\ {\frac {{\partial}{\phi}}{{\partial}x}})\
{\psi}$\ .\ \ \ \ \ (2.4)}
\end{center}
\par
Inserting the eqs. (2.3,4) into eq. (2.1), will be (remember that
$e^{i\ S}$ is a common factor):
\begin{center}
{$i\ {\hbar}\ ({\frac {{\partial}{\phi}}{{\partial}t}}\ +\ i\ {\phi}\
{\frac {{\partial}S}{{\partial}t}})\ =\ -\ {\frac {{\hbar}^{2}}{2\ m}}\
[{\frac {{\partial}^{2}{\phi}}{{\partial}x^{2}}}\ +$}
\end{center}
\begin{center}
{$+\ 2\ i\ {\frac {{\partial}{\phi}}{{\partial}x}}\ {\frac
{{\partial}S}{{\partial}x}}\ +\ i\ {\phi}\ {\frac
{{\partial}^{2}S}{{\partial}x^{2}}}\ -\ {\phi}\ ({\frac
{{\partial}S}{{\partial}x}})^{2}]\ +$}
\end{center}
\begin{center}
{$+\ {\Big {[}}\ V(x,\ t)\ +\ {\nu}\ {\Big {(}}\ [x\ -\ q(t)]\ [c\
{\hbar}\ ({\frac {{\partial}S}{{\partial}x}}\ -\ {\frac {i}{{\phi}}}\
{\frac {{\partial}{\phi}}{{\partial}x}})\ +$}
\end{center}
\begin{center}
{$+\ (1\ -\ c)\ <{\hat {p}}>]\ -\ {\frac
{1}{2}}\ i\ {\hbar}\ c\ {\Big {)}}\ {\Big {]}}\ {\phi}$\ .\ \ \ \ \
(2.5)}
\end{center}
\par
Taking the real and imaginary parts of eq. (2.5), we obtain (remember
that:\ $<\ {\hat {p}}\ >\ =\ m\ <\ {\hat {v}}_{qu}\ >\ =\ m\ <\ v_{qu}\
>$\ =\ real): 
\par
a)\ {\underline {imaginary part}}
\begin{center}
{${\frac {{\hbar}}{{\phi}}}\ {\frac {{\partial}{\phi}}{{\partial}t}}\
=\ -\ {\frac {{\hbar}^{2}}{2\ m}}\ ({\frac
{{\partial}^{2}S}{{\partial}x^{2}}}\ +\ 2\ {\frac {1}{{\phi}}}\ {\frac
{{\partial}{\phi}}{{\partial}x}}\ {\frac {{\partial}S}{{\partial}x}})\
-$}
\end{center}
\begin{center}
{$-\ {\nu}\ [x\ -\ q(t)]\ c\ {\frac {{\hbar}}{{\phi}}}\ {\frac
{{\partial}{\phi}}{{\partial}x}}\ -\ {\frac {{\nu}}{2}}\ {\hbar}\ c$\
,\ \ \ \ \ (2.6)}
\end{center}
\par
b)\ {\underline {real part}}
\begin{center}
{-\ ${\hbar}\ {\frac {{\partial}S}{{\partial}t}}\ =\ -\ {\frac
{{\hbar}^{2}}{2\ m}}\ [{\frac {1}{{\phi}}}\ {\frac
{{\partial}^{2}{\phi}}{{\partial}x^{2}}}\ -\ ({\frac
{{\partial}S}{{\partial}x}})^{2}]$\ +}
\end{center}
\begin{center}
{+\ ${\nu}\ [x\ -\ q(t)]\ c\ {\hbar}\ {\frac
{{\partial}S}{{\partial}x}}\ +\ V(x,\ t)\ +$}
\end{center}
\begin{center}
{$+\ {\nu}\ [x\ -\ q(t)]\ (1\ -\ c)\ m\ <v_{qu}>$\ .\ \ \ \ \ (2.7)}
\end{center}
\par
\vspace{0.2cm}
2.1 {\bf Dynamics of the S\"{u}ssmann-Hasse-Albrecht-Kostion-Nassar
Equation}
\par
Now, let us to study the dynamics of the
S\"{u}ssmann-Hasse-Albrecht-Kostin-Nassar equation. To do is let us
perform the following correspondences:$^{[9]}$
\begin{center}
{${\rho}(x,\ t)\ =\ {\phi}^{2}(x,\ t)$\ ,\ \ \ \ \ (2.8)\ \ \
(quantum mass density)}
\end{center}
\begin{center}
{$v_{qu}(x,\ t)\ =\ {\frac {{\hbar}}{m}}\ {\frac {{\partial}S(x,\
t)}{{\partial}x}}$\ ,\ \ \ \ \ (2.9)\ \ \ \ \ (quantum velocity)}
\end{center}
\begin{center}
{$V_{qu}(x,\ t)\ =\ -\ {\frac {{\hbar}^{2}}{2\ m}}\ {\frac {1}{{\sqrt
{{\rho}}}}}\ {\frac {{\partial}^{2}{\sqrt
{{\rho}}}}{{\partial}x^{2}}}\ =\ -\ {\frac {{\hbar}^{2}}{2\ m\
{\phi}}}\ {\frac {{\partial}^{2}{\phi}}{{\partial}x^{2}}}$\
.\ \ \ \ \ (2.10a,b)\ \ \ \ \ (Bohm quantum potential)}
\end{center}
\par
Putting the eqs. (2.8,9) into eq. (2.6) we get [remember that ${\frac
{{\partial}}{{\partial}v}}\ ({\ell}n\ u)\ =\ {\frac {1}{u}}\ {\frac
{{\partial}u}{{\partial}v}}$\ and\ ${\ell}n\ (u^{n})\ =\ n\ {\ell}n\
u$]:
\begin{center}
{${\frac {{\partial}}{{\partial}t}}\ (2\ {\ell}n\ {\phi})\ =\ -\ {\frac
{{\hbar}}{m}}\ [{\frac {{\partial}^{2}S}{{\partial}x^{2}}}\ +\ {\frac
{{\partial}S}{{\partial}x}}\ {\frac {{\partial}}{{\partial}x}}\ (2\
{\ell}n\ {\phi})]\ -$}
\end{center}
\begin{center}
{$-\ {\nu}\ [x\ -\ q(t)]\ c\ {\frac {{\partial}}{{\partial}x}}\ (2\
{\ell}n\ {\phi})\ -\ {\nu}\ c\ \ \ {\to}$}
\end{center}
\begin{center}
{${\frac {{\partial}}{{\partial}t}}\ ({\ell}n\ {\phi}^{2})\ =\ -\ {\frac
{{\hbar}}{m}}\ [{\frac {{\partial}^{2}S}{{\partial}x^{2}}}\ +\ {\frac
{{\partial}S}{{\partial}x}}\ {\frac {{\partial}}{{\partial}x}}\
({\ell}n\ {\phi}^{2})]\ -$}
\end{center}
\begin{center}
{$-\ {\nu}\ [x\ -\ q(t)]\ c\ {\frac {{\partial}}{{\partial}x}}\
({\ell}n\ {\phi}^{2})\ -\ {\nu}\ c\ \ \ {\to}$}
\end{center}
\begin{center}
{${\frac {{\partial}}{{\partial}t}}\ ({\ell}n\ {\rho})\ =\ -\ {\frac
{{\hbar}}{m}}\ [{\frac {{\partial}^{2}S}{{\partial}x^{2}}}\ +\ {\frac
{{\partial}S}{{\partial}x}}\ {\frac
{{\partial}}{{\partial}x}}\ ({\ell}n\ {\rho})]\ -$}
\end{center}
\begin{center}
{$-\ {\nu}\ [x\ -\ q(t)]\ c\ {\frac {{\partial}}{{\partial}x}}\
({\ell}n\ {\rho})\ -\ {\nu}\ c\ \ \ {\to}$}
\end{center}
\begin{center}
{${\frac {1}{{\rho}}}\ {\frac {{\partial}{\rho}}{{\partial}t}}\ =\ -\
{\frac {{\hbar}}{m}}\ ({\frac {{\partial}^{2}S}{{\partial}x^{2}}}\ +\
{\frac {{\partial}S}{{\partial}x}}\ {\frac {1}{{\rho}}}\ {\frac
{{\partial}{\rho}}{{\partial}x}})\ -$}
\end{center}
\begin{center}
{$-\ {\nu}\ [x\ -\ q(t)]\ c\ {\frac {1}{{\rho}}}\ {\frac
{{\partial}{\rho}}{{\partial}x}}\ -\ {\nu}\ c\ \ \ {\to}$}
\end{center}
\begin{center}
{${\frac {1}{{\rho}}}\ {\frac {{\partial}{\rho}}{{\partial}t}}\ +\
{\frac {{\partial}v_{qu}}{{\partial}x}}\ +\ {\frac {v_{qu}}{{\rho}}}\
{\frac {{\partial}{\rho}}{{\partial}x}}\ =\ -\ {\nu}\ c\ -\ {\nu}\ c\
[x\ -\ q(t)]\ {\frac {1}{{\rho}}}\ {\frac
{{\partial}{\rho}}{{\partial}x}}\ \ \ {\to}$}
\end{center}
\begin{center}
{${\frac {{\partial}{\rho}}{{\partial}t}}\ +\ {\frac {{\partial}({\rho}\
v_{qu})}{{\partial}x}}\ =$}
\end{center}
\begin{center}
{$=\ -\ {\nu}\ c\ {\rho}\ -\ {\nu}\ c\ [x\ -\ q(t)]\ {\frac
{{\partial}{\rho}}{{\partial}x}}$\ .\ \ \ \ \ (2.11)}
\end{center}
\par
We must note that the presence of the second member in expression (2.11),
indicates {\underline {descoerence}} of the considered physical system
represented by (2.1).
\par
Now, taking the eq. (2.7) and using the eqs. (2.9,10b), will be:
\par
\begin{center}
{$\ -\ {\hbar}\ {\frac {{\partial}S}{{\partial}t}}\ =\ -\ {\frac
{{\hbar}^{2}}{2\ m}}\ {\frac {1}{{\phi}}}\ {\frac
{{\partial}^{2}{\phi}}{{\partial}x^{2}}}\ +\ {\frac {1}{2}}\ m\ ({\frac
{{\hbar}}{m}}\ {\frac {{\partial}S}{{\partial}x}})^{2}$\ +}
\end{center}
\begin{center}
{$+\ m\ {\nu}\ [x\ -\ q(t)]\ c\ ({\frac {{\hbar}}{m}}\ {\frac
{{\partial}S}{{\partial}x}})\ +\ V(x,\ t)\ +$}
\end{center}
\begin{center}
{$+\ {\nu}\ [x\ -\ q(t)]\ (1\ -\ c)\ m\ <v_{qu}>\ \ \ {\to}$}
\end{center}
\begin{center}
{$-\ {\hbar}\ {\frac {{\partial}S}{{\partial}t}}\ =\
V_{qu}(x,\ t)\ +\ {\frac {1}{2}}\ m\ v_{qu}^{2}\ +\ m\ {\nu}\ [x\ -\
q(t)]\ c\ v_{qu}\ +\ V(x,\ t)\ +$}
\end{center}
\begin{center}
{$+\ {\nu}\ [x\ -\ q(t)]\ (1\ -\ c)\ m\ <v_{qu}>\ \ \ {\to}$}
\end{center} 
\begin{center}
{${\hbar}\ {\frac {{\partial}S}{{\partial}t}}\ +\ {\nu}\ m\ [x\ -\
q(t)]\ [c\ v_{qu}\ +\ (1\ -\ c)\ <v_{qu}>]\ +$}
\end{center}
\begin{center}
{$+\ {\frac {1}{2}}\ m\ v_{qu}^{2}\ +\ V_{qu}(x,\ t)\ +\ V(x,\ t)\ =\
0$\ . \ \ \ \ \ (2.12)}
\end{center}   
Considering that:
\begin{center}
{$<f(x,\ t)>\ =\ {\int}_{-\ {\infty}}^{+\ {\infty}}\ {\rho}(x,\ t)\
f(x,\ t) dx\ =\ g(t)$\ ,\ \ \ \ \ (2.13)}
\end{center}
where ${\rho}(x,\ t)$ is given by:$^{[10]}$
\begin{center}
{${\rho}\ (x,\ t) =\ [2{\pi}\ a^{2}(t)]^{-\ 1/2}\ exp\ {\Big {(}}\ -\
{\frac {[x\ -\ q(t)]^{2}}{2\ a^{2}(t)}}\ {\Big {)}}$,\ \ \ \ \ (2.14)}
\end{center}
where $a(t)$ and $q(t)\ =\ <\ x\ >$ are auxiliary functions of time, to be
determined in what follows;\ they represent the {\it width} and {\it center
of mass of wave packet}, respectively.
\par
Then, using the eqs. (2.13,14) and remembering that ${\int}_{-\
{\infty}}^{+\ {\infty}}\ exp (-\ z^{2})\ dz\ =\ {\sqrt {{\pi}}}$, will
be:
\begin{center}
{$<q(t)>\ =\ {\int}_{-\ {\infty}}^{+\ {\infty}}\ {\rho}(x,\ t)\
q(t) dx\ =\ q(t)$\ ,\ \ \ \ \ (2.15)}
\end{center}
\begin{center}
{${\frac {{\partial}<\ v_{vu}\ >}{{\partial}x}}\ =\ {\frac
{{\partial}}{{\partial}x}}\ {\int}_{-\ {\infty}}^{+\ {\infty}}\
{\rho}(x,\ t)\ v_{qu}\ (x,\ t) dx\ =\ {\frac
{{\partial}g(t)}{{\partial}x}}\ =\ 0$\ ,\ \ \ \ \ (2.16)}
\end{center}
and (remember that $x$ and $t$ are variables independents):
\begin{center}
{${\frac {{\partial}q(t)}{{\partial}x}}\ =\ 0$\ .\ \ \ \ \ (2.17)}
\end{center}
\par
Defining:$^{[11]}$
\begin{center}
{${\vartheta}_{qnc}\ =\ v_{qu}\ +\ {\nu}\ c\ [x\ -\ q(t)]$\ ,\ \ \ \ \
(2.18)}
\end{center}
the {\bf quantum velocity non-conservative}, we have [by using the eq. (2.17)]:
\begin{center}
{${\frac {{\partial}}{{\partial}x}}\ ({\rho}\
{\vartheta}_{qnc})\ =\ {\frac {{\partial}}{{\partial}x}}\ {\big {[}}\
{\rho}\ {\big {(}}\ v_{qu}\ +\ {\nu}\ c\ [x\ -\ q(t)]\ {\big {)}}\
{\big {]}}\ =$}
\end{center}
\begin{center}
{$=\ {\frac {{\partial}({\rho}\ v_{qu})}{{\partial}x}}\ +\ {\frac
{{\partial}}{{\partial}x}}\ {\big {(}}\ {\rho}\ {\nu}\ c\ [x\ -\ q(t)]\
{\big {)}}\ =\ {\frac {{\partial}({\rho}\ v_{qu})}{{\partial}x}}\ +$}
\end{center}
\begin{center}
{$+\ {\nu}\ c\ [x\ -\ q(t)]\ {\frac {{\partial}{\rho}}{{\partial}x}}\ +\
{\rho}\ {\nu}\ c\ {\frac {{\partial}}{{\partial}x}}\ [x\ -\ q(t)]\ =$}
\end{center}
\begin{center}
{$=\ {\frac {{\partial}({\rho}\ v_{qu})}{{\partial}x}}\ +\ {\nu}\ c\
[x\ -\ q(t)]\ {\frac {{\partial}{\rho}}{{\partial}x}}\ +\ {\nu}\ c\
{\rho}\ \ \ {\to}$}
\end{center}
\begin{center}
{$-\ {\nu}\ c\ {\rho}\ -\ {\nu}\ c\ [x\ -\ q(t)]\ {\frac
{{\partial}{\rho}}{{\partial}x}}\ =$}
\end{center} 
\begin{center}
{$=\ {\frac {{\partial}({\rho}\ v_{qu})}{{\partial}x}}\ -\ {\frac
{{\partial}({\rho}\ {\vartheta}_{qnc})}{{\partial}x}}$\ .\
\ \ \ \ (2.19)}
\end{center}
\par
Insering the eq. (2.19) into eq. (2.12), results: 
\begin{center}
{${\frac {{\partial}{\rho}}{{\partial}t}}\ +\ {\frac {{\partial}({\rho}\
{\vartheta}_{qnc})}{{\partial}x}}\ =\ 0$\ .\ \ \ \ \ (2.20)}
\end{center}
\par
The\ eq. (2.20) indicates that, considering the {\bf quantum
velocity non-conservative} (${\vartheta}_{qnc}$), there are {\underline
{coerence}} of the considered physical system represented by (2.1).
\par
Now, differentiating the eq. (2.7) with respect $x$, and using the eqs.
(2.9,10b,16-18), we have (remember that $x$\ and $t$\ are variables
independents):
\begin{center}
{$-\ {\hbar}\ {\frac {{\partial}^{2}S}{{\partial}x\ {\partial}t}}\ =\
-\ {\frac {{\hbar}^{2}}{2\ m}}\ {\frac {{\partial}}{{\partial}x}}\
[{\frac {1}{{\phi}}}\ {\frac {{\partial}^{2}{\phi}}{{\partial}x^{2}}}\
-\ ({\frac {{\partial}S}{{\partial}x}})^{2}]\ +$}
\end{center}
\begin{center}
{$+\ {\frac {{\partial}}{{\partial}x}}\ {\Big {(}}\ {\nu}\ [x\ -\ q(t)]\ c\
{\hbar}\ {\frac {{\partial}S}{{\partial}x}}\ +$}
\end{center}
\begin{center}
{$+\ {\nu}\ [x\ -\ q(t)]\ (1\ -\ c)\ m\ <v_{qu}>\ +\ V(x,\ t)\ {\Big
{)}}\ \ \ {\to}$}
\end{center}
\begin{center}
{${\frac {{\partial}}{{\partial}t}}\ ({\frac {{\hbar}}{m}}\ {\frac
{{\partial}S}{{\partial}x}})\ =\ {\frac {1}{m}}\ {\frac
{{\partial}}{{\partial}x}}\ ({\frac {{\hbar}^{2}}{2\ m}}\ {\frac
{1}{{\phi}}}\ {\frac {{\partial}^{2}{\phi}}{{\partial}x^{2}}})\ -\
{\frac {1}{2}}\ {\frac {{\partial}}{{\partial}x}}\ ({\frac
{{\hbar}}{m}}\ {\frac {{\partial}S}{{\partial}x}})^{2}\ -$}
\end{center}
\begin{center}
{$-\ {\frac {{\partial}}{{\partial}x}}\ {\Big {(}}\ {\nu}\ [x\ -\ q(t)]\ c\
({\frac {{\hbar}}{m}}\ {\frac {{\partial}S}{{\partial}x}})\ {\Big {)}}\ -$}
\end{center}
\begin{center}
{$-\ {\frac {{\partial}}{{\partial}x}}\ {\Big {(}}\ {\nu}\ [x\ -\
q(t)]\ (1\ -\ c)\ <v_{qu}>\ {\Big {)}}\ -\ {\frac {1}{m}}\ {\frac
{{\partial}V}{{\partial}x}}\ \ \ {\to}$}
\end{center}
\begin{center}
{${\frac {{\partial}v_{qu}}{{\partial}t}}\ =\ -\ {\frac {1}{m}}\ {\frac
{{\partial}V_{qu}}{{\partial}x}}\ -\ {\frac {1}{2}}\ {\frac
{{\partial}}{{\partial}x}}\ (v_{qu}^{2})\ -\ {\frac
{{\partial}}{{\partial}x}}\ {\Big {(}}\ {\nu}\ c\ v_{qu}\ [x\ -\ q(t)]\
{\Big {)}}\ -$}
\end{center}
\begin{center}
{$-\ {\frac {{\partial}}{{\partial}x}}\ {\Big {(}}\ {\nu}\ [x\ -\
q(t)]\ (1\ -\ c)\ <v_{qu}>\ {\Big {)}}\ -\ {\frac {1}{m}}\ {\frac
{{\partial}V}{{\partial}x}}\ \ \ {\to}$}
\end{center}
\begin{center}
{${\frac {{\partial}v_{qu}}{{\partial}t}}\ +\ v_{qu}\ {\frac
{{\partial}v_{qu}}{{\partial}x}}\ +\ v_{qu}\ {\frac
{{\partial}}{{\partial}x}}\ {\Big {(}}\ {\nu}\ c\ [x\ -\ q(t)]\ {\Big
{)}}\ +$}
\end{center}
\begin{center}
{$+\ {\nu}\ c\ [x\ -\ q(t)]\ {\frac {{\partial}v_{qu}}{{\partial}x}}\ +$}
\end{center}
\begin{center}
{$+\ {\nu}\ (1\ -\ c)\ <v_{qu}>\ {\frac {{\partial}}{{\partial}x}}\ [x\ -\
q(t)]\ =$}
\end{center}
\begin{center}
{$=\ -\ {\frac {1}{m}}\ {\frac {{\partial}}{{\partial}x}}\ (V\ +\
V_{qu})\ \ \ {\to}$}
\end{center}
\begin{center}
{${\frac {{\partial}v_{qu}}{{\partial}t}}\ +\ {\frac
{{\partial}v_{qu}}{{\partial}x}}\ {\Big {(}}\ v_{qu}\ +\ {\nu}\ c\ [x\
-\ q(t)]\ {\Big {)}}\ +$}
\end{center}
\begin{center}
{$+\ [{\nu}\ v_{qu}\ c\ +\ {\nu}\ (1\ -\ c)\ <v_{qu}>]\ =$}
\end{center}
\begin{center}
{$=\ -\ {\frac {1}{m}}\ {\frac {{\partial}}{{\partial}x}}\ (V\ +\
V_{qu})\ \ \ {\to}$}
\end{center}
\begin{center}
{${\frac {{\partial}v_{qu}}{{\partial}t}}\ +\ {\vartheta}_{qnc}\ {\frac
{{\partial}v_{qu}}{{\partial}x}}\ +\ {\frac {1}{m}}\ {\frac
{{\partial}}{{\partial}x}}\ (V\ +\ V_{qu})\ =$}
\end{center}
\begin{center}
{$=\ -\ {\nu}\ [c\ v_{qu}\ +\ (1\ -\ c)\ <v_{qu}>]$\ .\ \ \ \ \ (2.21)}
\end{center}
\par
Taking the eqs. (2.15,17,18) and considering that $< >$ is a linear
operation, results [remember that $<x>$\ =\ q(t)]: 
\begin{center}
{$<{\vartheta}_{qnc}>\ =\ <v_{qu}>\ +\ {\nu}\ c\ (<x>\ -\ <q(t)>)\ \ \ {\to}$}
\end{center}
\begin{center}
{$<{\vartheta}_{qnc}>\ =\ <v_{qu}>$\ .\ \ \ \ \ (2.22)}
\end{center}
\par
Differentiating the eq. (2.18) with respect $t$, considering that
$<v_{qu}>\ =\ {\frac {{\partial}<x>}{{\partial}t}}\ =\ {\dot {q}}(t)$,
$<x>$\ =\ q(t) and the eq. (2.22) (remember that $x$ and $t$ are
variables independents), we have:
\begin{center}
{${\frac {{\partial}{\vartheta}_{qnc}}{{\partial}t}}\ =\ {\frac
{{\partial}v_{qu}}{{\partial}t}}\ +\ {\nu}\ c\ ({\frac
{{\partial}x}{{\partial}t}}\ -\ {\frac {{\partial}<x>}{{\partial}t}})\ \
\ {\to}$}
\end{center}
\begin{center}
{${\frac {{\partial}v_{qu}}{{\partial}t}}\ =\ {\frac
{{\partial}{\vartheta}_{qnc}}{{\partial}t}}\ +\ {\nu}\ c\
<{\vartheta}_{qnc}>$\ .\ \ \ \ \ (2.23)}
\end{center}
\par
Now, differentiating the eq. (2.18) with respect $x$ and considering the eq.
(2.17), results:
\begin{center}
{${\frac {{\partial}{\vartheta}_{qnc}}{{\partial}x}}\ =\ {\frac
{{\partial}v_{qu}}{{\partial}x}}\ +\ {\nu}\ c\ =\ \ \ {\to}\ \ \
{\frac {{\partial}v_{qu}}{{\partial}x}}\ =\ {\frac
{{\partial}{\vartheta}_{qnc}}{{\partial}x}}\ -\ {\nu}\ c$\ .\ \ \ \ \
(2.24)}
\end{center} 
\par
Substituting the eqs. (2.22-24) into eq. (2.21), we have:
\begin{center}
{${\frac {{\partial}{\vartheta}_{qnc}}{{\partial}t}}\ +\ {\nu}\ c\
<{\vartheta}_{qnc}>\ +$}
\end{center}
\begin{center}
{$+\ {\vartheta}_{qnc}\ ({\frac
{{\partial}{\vartheta}_{qnc}}{{\partial}x}}\ -\ {\nu}\ c)\ +\ {\frac
{1}{m}}\ {\frac {{\partial}}{{\partial}x}}\ (V\ +\ V_{qu})\ =$}
\end{center}
\begin{center}
{$=\ -\ {\nu}\ [c\ v_{qu}\ +\ (1\ -\ c)\ <{\vartheta}_{qnc}>]\ =$}
\end{center}
\begin{center}
{$=\ -\ {\nu}\ c\ v_{qu}\ -\ {\nu}\ <{\vartheta}_{qnc}>\ +\ {\nu}\ c\
<{\vartheta}_{qnc}>\ \ \ {\to}$}
\end{center} 
\begin{center}
{${\frac {{\partial}{\vartheta}_{qnc}}{{\partial}t}}\ +\ {\nu}\ c\
<{\vartheta}_{qnc}>\ +$}
\end{center}
\begin{center}
{$+\ {\vartheta}_{qnc}\ {\frac
{{\partial}{\vartheta}_{qnc}}{{\partial}x}}\ -\ {\vartheta}_{qnc}\
{\nu}\ c\ +\ {\frac {1}{m}}\ {\frac {{\partial}}{{\partial}x}}\ (V\ +\
V_{qu})\ =$}
\end{center}
\begin{center}
{$=\ -\ {\nu}\ c\ v_{qu}\ -\ {\nu}\ <{\vartheta}_{qnc}>\ +\ {\nu}\ c\
<{\vartheta}_{qnc}>\ \ \ {\to}$}
\end{center}
\begin{center}
{${\frac {{\partial}{\vartheta}_{qnc}}{{\partial}t}}\ +\
{\vartheta}_{qnc}\ {\frac {{\partial}{\vartheta}_{qnc}}{{\partial}x}}\ +\
{\frac {1}{m}}\ {\frac {{\partial}}{{\partial}x}}\ (V\ +\ V_{qu})\ =$}
\end{center}
\begin{center}
{$=\ -\ {\nu}\ [c\ (v_{qu}\ -\ {\vartheta}_{qnc})\ +\
<{\vartheta}_{qnc}>]$\ .\ \ \ \ \ (2.25)}
\end{center}
\par
Considering the "substantive differentiation" (local plus convective)
or "hidrodynamic differention":\ $d/dt\ =\ {\partial}/{\partial}t\ +
{\vartheta}_{qnc}\ {\partial}/{\partial}x$ and that ${\vartheta}_{qnc}\
=\ dx_{qnc}/dt$, the eq. (2.25) could be written as:$^{[9]}$
\begin{center}
{$m\ d^{2}x_{qnc}/dt^{2}\ =\ -\ {\nu}\ m\ [c\ (v_{qu}\ -\
{\vartheta}_{qnc})\ +\ <\ {\vartheta}_{qnc}>]\ -\ {\frac
{{\partial}}{{\partial}x}}\ (V\ +\ V_{qu})$\ ,\ \ \ \ \ (2.26)}
\end{center}
that has a form of the $Second\ Newton\ Law$.
\vspace{0.2cm}
\par
3.\ {\bf The Quantum Wave Packet of the Linearized
S\"{u}ssmann-Hasse-Albrecht-Kostin-Nassar Equation along a Classical
Trajetory}
\vspace{0.2cm}
\par
In order to find the quantum wave packet of the
S\"{u}ssmann-Hasse-Albrecht-Koston-Nassar equation, let us considerer
the eq. (2.13):
\par
\begin{center}
{${\rho}\ (x,\ t) =\ [2{\pi}\ a^{2}(t)]^{-\ 1/2}\ exp\ {\Big {(}}\ -\
{\frac {[x\ -\ q(t)]^{2}}{2\ a^{2}(t)}}\ {\Big {)}}$.\ \ \ \ \ (3.1)}
\end{center}
\par
Using the eq. (3.1), the integration of the eq. (2.20), is given
by:$^{[1]}$ 
\begin{center}
{${\vartheta}_{qu}(x,\ t)\ =\ {\frac {{\dot {a}}(t)}{a(t)}}\ [x\ -\
q(t)]\ +\ {\dot {q}}(t)$\ ,\ \ \ \ \ (3.2a)}
\end{center}
where ${\dot {q}}(t)\ =\ {\frac {d<x>}{dt}}$.
\par 
Now, using the eq. (3.2a) into eq. (2.18), we have:
\begin{center}
{$v_{qu}\ (x,\ t)\ =\ [{\frac {{\dot
{a}}(t)}{a(t)}}\ -\ {\nu}\ c]\ [x\ -\ q(t)]\ +\ {\dot {q}}(t)$\
.\ \ \ \ \ (3.2b)}
\end{center}
\par
To obtain the quantum wave packet of the linear
S\"{u}ssmann-Hasse-Albrecht-Kostin-Nassar equation along a classical
trajetory given by (2.1), let us expand the functions $S(x,\ t)$,\ \
$V(x,\ t)$ and $V_{qu}(x,\ t)$ around of $q(t)$ up to second Taylor
order.\ In this way we have:
\begin{center}
{$S(x,\ t)\ =\ S[q(t),\ t]\ +\ S'[q(t),\ t]\ [x\ -\ q(t)]\ +\ {\frac
{S''[q(t),\ t]}{2}}\ [x\ -\ q(t)]^{2}$\ ,\ \ \ \ \ (3.3)}
\end{center}
\begin{center}
{$V(x,\ t)\ =\ V[q(t),\ t]\ +\ V'[q(t),\ t]\ [x\ -\
q(t)]\ +\ {\frac {V''[q(t),\ t]}{2}}\ [x\ -\ q(t)]^{2}$\ .\ \ \ \ \ (3.4)}
\end{center}
\begin{center}
{$V_{qu}(x,\ t)\ =\ V_{qu}[q(t),\ t]\ +\ V_{qu}'[q(t),\ t]\ [x\ -\
q(t)]\ +\ {\frac {V_{qu}''[q(t),\ t]}{2}}\ [x\ -\ q(t)]^{2}$\ .\ \ \ \
\ (3.5)}
\end{center}
\par
Differentiating (3.3) in the variable $x$, multiplying
the result by ${\frac {{\hbar}}{m}}$, using the eqs. (2.9) and
(3.2b), taking into account the polynomial identity property and also
considering the second Taylor order, we obtain:
\begin{center}
{${\frac {{\hbar}}{m}}\ {\frac {{\partial}S(x,\ t)}{{\partial}x}}\ =\
{\frac {{\hbar}}{m}}\ {\Big {(}}\ S'[q(t),\ t]\ +\ S''[q(t),\ t]\ [x\
-\ q(t)]\ {\Big {)}}\ =$}
\end{center}
\begin{center}
{$=\ v_{qu}(x,\ t)\ =\ [{\frac {{\dot
{a}}(t)}{a(t)}}\ -\ {\nu}\ c]\ [x\ -\ q(t)]\ +\ {\dot {q}}(t)\
\ \ \ {\to}$}
\end{center}
\begin{center}
{$S'[q(t),\ t]\ =\ {\frac {m\ {\dot {q}}(t)}{{\hbar}}}\ ,\ \ \
S''[q(t),\ t]\ =\ {\frac {m}{{\hbar}}}\ [{\frac {{\dot
{a}}(t)}{a(t)}}\ -\ {\nu}\ c]$\ ,\ \ \ \ \ (3.6a,b)}
\end{center}
\par
Substituting (3.6a,b) into (3.3), results:
\begin{center}
{$S(x,\ t)\ =\ S_{o}(t)\ +\ {\frac {m\ {\dot {q}}(t)}{{\hbar}}}\ [x\ -\
q(t)]\ +\ {\frac {m}{2\ {\hbar}}}\ [{\frac {{\dot
{a}}(t)}{a(t)}}\ -\ {\nu}\ c]\ [x\
-\ q(t)]^{2}$\ ,\ \ \ \ \ (3.7a)}
\end{center}
where:
\begin{center}
{$S_{o}(t)\ {\equiv}\ S[q(t),\ t]$\ ,\ \ \ \ \ (3.7b)}
\end{center}
are the classical actions.
\par
\par
Differentiating the (3.7a) with respect to $t$, we obtain
(remembering that ${\frac {{\partial}x}{{\partial}t}}$\ =\ 0):
\begin{center}
{${\frac {{\partial}S\ (x,\ t)}{{\partial}t}}\ =\ {\dot {S}}_{o}(t)\ +\
{\frac {{\partial}}{{\partial}t}}\ {\big {[}}\ {\frac {m\ {\dot
{q}}(t)}{{\hbar}}}\ [x\ -\ q(t)]\ {\big {]}}\ +\ {\frac
{{\partial}}{{\partial}t}}\ {\big {[}}\ {\frac {m}{2\ {\hbar}}}\ {\big
{(}}\ [{\frac {{\dot {a}}(t)}{a(t)}}\ -\ {\nu}\ c]\ {\big {)}}\ [x\ -\
q(t)]^{2}\ {\big {]}}\ \ \ {\to}$}
\end{center}
\begin{center}
{${\frac {{\partial}S\ (x,\ t)}{{\partial}t}}\ =\ {\dot {S}}_{o}(t)\ +\
{\frac {m\ {\ddot {q}}(t)}{{\hbar}}}\ [x\ -\ q(t)]\ -\ {\frac {m\ {\dot
{q}}^{2}(t)}{{\hbar}}}\ +$}
\end{center}
\begin{center}
{+\ ${\frac {m}{2\ {\hbar}}}\ [{\frac {{\ddot
{a}}(t)}{a(t)}}\ -\ {\frac {{\dot
{a}}^{2}(t)}{a^{2}(t)}}]\ [x\ -\ q(t)]^{2}\ -\ {\frac {m\
{\dot {q}}(t)}{{\hbar}}}\ {\big {[}}\ {\frac {{\dot
{a}}(t)}{a(t)}}\ -\ {\nu}\ c\ {\big {]}}\ [x\ -\ q(t)]$\ .\ \
\ \ \ (3.8)}
\end{center}
\par
Considering the eqs. (2.8) and (3.1), let us write $V_{qu}$ given by
(2.10a,b) in terms of potencies of $[x\ -\ q(t)]$. Before, we calculate
the following derivations:
\begin{center}
{${\frac {{\partial}{\phi}\ (x,\ t)}{{\partial}x}}\ =\ {\frac
{{\partial}}{{\partial}x}}\ {\big {(}}\ [2\ {\pi}\ a^{2}(t)]^{-\
1/4}\ e^{-\ {\frac {[x\ -\ q(t)]^{2}}{4\ a^{2}(t)}}}\ {\big
{)}}\ =$}
\end{center}
\begin{center}
{$=\ [2\ {\pi}\ a^{2}(t)]^{-\ 1/4}\ e^{-\ {\frac {[x\ -\
q(t)]^{2}}{4\ a^{2}(t)}}} {\frac {{\partial}}{{\partial}x}}\
{\big {(}}\ -\ {\frac {[x\ -\ q(t)]^{2}}{4\ a^{2}(t)}}\ {\big
{)}}\ \ \ {\to}$}
\end{center}
\begin{center}
{${\frac {{\partial}{\phi}\ (x,\ t)}{{\partial}x}}\ =\ -\ [2\ {\pi}\
a^{2}(t)]^{-\ 1/4}\ e^{-\ {\frac {[x\ -\ q(t)]^{2}}{4\
^{2}(t)}}}\ {\frac {[x\ -\ q(t)]}{2\ a^{2}(t)}}$\ ,}
\end{center}
\begin{center}
{${\frac {{\partial}^{2}{\phi}\ (x,\ t)}{{\partial}x^{2}}}\ =\ {\frac
{{\partial}}{{\partial}x}}\ {\big {(}}\ -\ [2\ {\pi}\
a^{2}(t)]^{-\ 1/4}\ e^{-\ {\frac {[x\ -\ q(t)]^{2}}{4\
a^{2}(t)}}}\ {\frac {[x\ -\ q(t)]}{2\ a^{2}(t)}}\ {\big
{)}}$\ =}
\end{center}
\begin{center}
{$\ =\ -\ [2\ {\pi}\ a^{2}(t)]^{-\ 1/4}\ e^{-\ {\frac {[x\ -\
q(t)]^{2}}{4\ a^{2}(t)}}}\ {\frac {{\partial}}{{\partial}x}}\
{\big {(}}\ {\frac {[x\ -\ q(t)]}{2\ a^{2}(t)}}\ {\big {)}}\ -$}
\end{center}
\begin{center}
{$-\ [2\ {\pi}\ a^{2}(t)]^{-\ 1/4}\ e^{-\ {\frac {[x\ -\
q(t)]^{2}}{4\ a^{2}(t)}}}\ {\frac {[x\ -\ q(t)]}{2\ a^{2}(t)}}\ {\frac
{{\partial}}{{\partial}x}}\ {\big {(}}\ -\ {\frac {[x\ -\ q(t)]^{2}}{4\
a^{2}(t)}}\ {\big {)}}\ \ \ {\to}$}
\end{center}
\begin{center}
{${\frac {{\partial}^{2}{\phi}\ (x,\ t)}{{\partial}x^{2}}}\ =\ -\ [2\
{\pi}\ a^{2}(t)]^{-\ 1/4}\ e^{-\ {\frac {[x\ -\ q(t)]^{2}}{4\
a^{2}(t)}}}\ {\frac {1}{2\ a^{2}(t)}}\ +\ [2\ {\pi}\ a^{2}(t)]^{-\
1/4}\ e^{-\ {\frac {[x\ -\ q(t)]^{2}}{4\ a^{2}(t)}}}\ {\frac {[x\ -\
q(t)]^{2}}{4\ a^{4}(t)}}$\ =}
\end{center}
\begin{center}
{$=\ -\ {\phi}\ (x,\ t)\ {\frac {1}{2\ a^{2}(t)}}\ +\ {\phi}\ (x,\ t)\
{\frac {[x\ -\ q(t)]^{2}}{4\ a^{4}(t)}}\ \ \ {\to}$}
\end{center}
\begin{center}
{${\frac {1}{{\phi}\ (x,\ t)}}\ {\frac {{\partial}^{2}{\phi}\ (x,\
t)}{{\partial}x^{2}}}\ =\ {\frac {[x\ -\ q(t)]^{2}}{4\ a^{4}(t)}}\ -\
{\frac {1}{2\ a^{2}(t)}}$\ .\ \ \ \ \ (3.9)}
\end{center}
\par
Substituting (3.9) into (2.10b) and taking into account (3.4), results:
\begin{center}
{$V_{qu}(x,\ t)\ =\ {\frac {{\hbar}^{2}}{4\ m\ a^{2}(t)}}\ -\ {\frac
{{\hbar}^{2}}{8\ m\ a^{4}(t)}}\ [x\ -\ q(t)]^{2}$\ .\ \ \ \ \ (3.10)}
\end{center}
\begin{center}
{$\ V_{qu}[q(t),\ t]\ =\ {\frac {{\hbar}^{2}}{4\ m\ a^{2}(t)}}$\ ,\ \ \
\ \ (3.11a)}
\end{center}
\begin{center}
{$\  V_{qu}'[q(t),\ t]\ =\ 0, \ \ \ V_{qu}''[q(t),\ t]\ =\ -\ {\frac
{{\hbar}^{2}}{4\ m\ a^{4}(t)}}$\ .\ \ \ \ \ (3.11b,c)}
\end{center}
\par
Inserting the eqs. (3.2b,3,4) and (3.7a,8,10), into (2.12), we obtain
[remem\-bering that $S_{o}(t)$, $a(t)$, $q(t)$ and $<v_{qu}>\ =\ {\frac
{d<x>}{dt}}\ =\ {\dot {q}}(t)$]:
\begin{center}
{${\hbar}\ {\frac {{\partial}S}{{\partial}t}}\ +\ {\nu}\ m\ [x\ -\
q(t)]\ [c\ v_{qu}\ +\ (1\ -\ c)\ <v_{qu}>]\ +\ {\frac {1}{2}}\ m\
v_{qu}^{2}\ +\ V\ +\ V_{qu}\ =$}
\end{center}
\begin{center}
{$=\ {\hbar}\ {\dot {S}}_{o}(t)\ +\ m\ {\ddot
{q}}(t)\ [x\ -\ q(t)]\ -\ m\ {\dot
{q}}^{2}(t)\ +\ {\frac {m}{2}}\ {\Big {[}}\ {\frac
{{\ddot {a}}(t)}{a(t)}}\ -\ {\frac {{\dot
{a}}^{2}(t)}{a^{2}(t)}}\ {\Big {]}}\ [x\ -\ q(t)]^{2}\ -$}
\end{center}
\begin{center}
{$-\ m\ {\dot {q}}(t)\ [{\frac {{\dot {a}}(t)}{a(t)}}\ -\ {\nu}\ c]\
[x\ -\ q(t)]\ +\ {\nu}\ m\ [x\ -\ q(t)]\ {\times}$}
\end{center}
\begin{center}
{${\times}\ {\Big {[}}\ c\ {\Big {(}}\ [{\frac {{\dot {a}}(t)}{a(t)}}\
-\ {\nu}\ c]\ [x\ -\ q(t)]\ +\ {\dot {q}}(t)\ {\Big {)}}\ +\ (1\ -\ c)\
{\dot {q}}(t)\ {\Big {]}}\ +$}
\end{center}
\begin{center}
{$+\ {\frac {m}{2}}\ [{\frac{{\dot {a}}(t)}{a(t)}}\ -\
{\nu}\ c]^{2}\ [x\ -\ q(t)]^{2}\ +\ m\ {\dot {q}}(t)\ [{\frac{{\dot
{a}}(t)}{a(t)}}\ -\ {\nu}\ c]\ [x\ -\ q(t)]\ +\ {\frac {m\
{\dot {q}}^{2}(t)}{2}}\ +$}
\end{center}
\begin{center}
{$+\ V[q(t),\ t]\ +\ V'[q(t),\ t]\ [x\ -\ q(t)]\ +\ {\frac {1}{2}}\
V''[q(t),\ t]\ [x\ -\ q(t)]^{2}$\ +}
\end{center}
\begin{center}
{$+\ {\frac {{\hbar}^{2}}{4\ m\ a^{2}(t)}}\ -\ {\frac
{{\hbar}^{2}}{8\ m\ a^{4}(t)}}\ [x\ -\ q(t)]^{2}\ =\ $}
\end{center}
\begin{center}
{$=\ {\hbar}\ {\dot {S}}_{o}(t)\ +\ m\ {\ddot
{q}}(t)\ [x\ -\ q(t)]\ -\ m\ {\dot
{q}}^{2}(t)\ +\ {\frac {m}{2}}\ {\Big {[}}\ {\frac
{{\ddot {a}}(t)}{a(t)}}\ -\ {\frac {{\dot
{a}}^{2}(t)}{a^{2}(t)}}\ {\Big {]}}\ [x\ -\ q(t)]^{2}\ -$}
\end{center}
\begin{center}
{$-\ m\ {\dot {q}}(t)\ [{\frac {{\dot {a}}(t)}{a(t)}}\ -\ {\nu}\ c]\
[x\ -\ q(t)]\ +\ {\nu}\ m\ c\ [{\frac {{\dot {a}}(t)}{a(t)}}\ -\ {\nu}\
c]\ [x\ -\ q(t)]^{2}\ +$}
\end{center}
\begin{center}
{$+\  {\nu}\ m\ c\ {\dot {q}}(t)\ [x\ -\ q(t)]\ +\ {\nu}\
m\ (1\ -\ c)\ {\dot {q}}(t)\ {\Big {)}}\ +$}
\end{center}
\begin{center}
{$+\ {\frac {m}{2}}\ [{\frac{{\dot {a}}(t)}{a(t)}}\ -\
{\nu}\ c]^{2}\ [x\ -\ q(t)]^{2}\ +\ m\ {\dot {q}}(t)\ [{\frac{{\dot
{a}}(t)}{a(t)}}\ -\ {\nu}\ c]\ [x\ -\ q(t)]\ +\ {\frac {m\
{\dot {q}}^{2}(t)}{2}}\ +$}
\end{center}
\begin{center}
{$+\ V[q(t),\ t]\ +\ V'[q(t),\ t]\ [x\ -\ q(t)]\ +\ {\frac {1}{2}}\
V''[q(t),\ t]\ [x\ -\ q(t)]^{2}$\ +}
\end{center}
\begin{center}
{$+\ {\frac {{\hbar}^{2}}{4\ m\ a^{2}(t)}}\ -\ {\frac
{{\hbar}^{2}}{8\ m\ a^{4}(t)}}\ [x\ -\ q(t)]^{2}\ =\ 0$.\ \ \ \ \ (3.12)}
\end{center}
\par
Expanding the eq. (3.12) in potencies of $[x\ -\ q(t)]$, we obtain
(remember that $[x\ -\ q(t)]^{o}\ =\ 1$):
\begin{center}
{${\Big {(}}\ {\hbar}\ {\dot {S}}_{o}(t)\ -\ {\frac {m\
{\dot {q}}^{2}(t)}{2}}\ +\ {\nu}\ m\ (1\ - c)\ {\dot
{q}}(t)\ +\ V[q(t),\ t]\ +\ {\frac {{\hbar}^{2}}{4\ m\ a^{2}(t)}}\
{\Big {)}}\ [x\ -\ q(t)]^{o}\ +$}
\end{center}
\begin{center}
{$+\ {\Big {(}}\ m\ {\ddot {q}}(t)\ +\ {\nu}\ m\ c\ {\dot {q}}(t)\ +\
V'[q(t),\ t]\ {\Big {)}}\ [x\ -\ q(t)]$\ +}
\end{center}
\begin{center}
{$+\ {\Big {(}}\ {\frac {m}{2}}\ {\frac {{\ddot {a}}(t)}{a(t)}}\ -\ {\frac
{m\ {\nu}^{2}\ c^{2}}{2}}\ +\ {\frac {1}{2}}\ V"[q(t),\ t]\ -\ {\frac
{{\hbar}^{2}}{8\ m\ a^{4}(t)}}\ {\Big {)}}\ [x\ -\ q(t)]^{2}\ =\ 0$\ .\
\ \ \ \ (3.13)}
\end{center}
\par
As (3.13) is an identically null polynomium, all coefficients of the
potencies must be all equal to zero, that is:
\begin{center}
{${\dot {S}}_{o}(t)\ =\ {\frac {1}{{\hbar}}} {\Big {[}}\ {\frac
{1}{2}}\ m\ {\dot {q}}^{2}(t)\ -\ {\nu}\ m\ (1\ -\ c)\ {\dot {q}}(t)\
-\ V[q(t),\ t]\ -\ {\frac {{\hbar}^{2}}{4\ m\ a^{2}(t)}}\ {\Big {]}}$\
,\ \ \ \ \ (3.14)}
\end{center}
\begin{center}
{${\ddot {q}}(t)\ +\ {\nu}\ c\ {\dot {q}}(t)\ +\ {\frac {1}{m}}\
V'[q(t),\ t]\ =\ 0$,\ \ \ \ \ (3.15)}
\end{center}
\begin{center}
{${\ddot {a}}(t)\ +\ a(t)\ {\Big {(}}\ {\frac {1}{m}}\ V"[q(t),\ t]\
-\ {\nu}^{2}\ c^{2}\ {\Big {)}}\ =\ {\frac {{\hbar}^{2}}{4\
m^{2}\ a^{3}(t)}}$\ .\ \ \ \ \ (3.16)}
\end{center}
\par
Assuming that the following initial conditions are obeyed:
\begin{center}
{$q(0)\ =\ x_{o}\ ,\ \ \ {\dot {q}}(0)\ =\ v_{o}\ ,\ \ \ a(0)\
=\ a_{o}\ ,\ \ \ {\dot {a}}(0)\ =\ b_{o}$\ ,\ \ \ \ \ \ (3.17a-d)}
\end{center}
and that:
\begin{center}
{$S_{o}(0)\ =\ {\frac {m\ v_{o}\ x_{o}}{{\hbar}}}$\ ,\ \ \ \ \ (3.18)}
\end{center}
the integration of (3.14) gives:
\begin{center}
{$S_{o}(t)\ =\ {\frac {1}{{\hbar}}}\ {\int}_{o}^{t}\ dt'\ {\Big {[}}\
{\frac {1}{2}}\ m\ {\dot {q}}^{2}(t')\ -\ {\nu}\ m\ (1\ -\ c)\ {\dot
{q}}(t')\ -$}
\end{center}
\begin{center}
{$-\ V[q(t'),\ t']\ -\ {\frac {{\hbar}^{2}}{4\ m\ a^{2}(t')}}\ {\Big
{]}}\ +\ {\frac {m\ v_{o}\ x_{o}}{{\hbar}}}\ $.\ \ \ \
\ (3.19)}
\end{center}
\par
Taking the eq. (3.19) in the eq. (3.7a) results:
\begin{center}
{$S(x,\ t)\ =\ {\frac {1}{{\hbar}}}\ {\int}_{o}^{t}\ dt'\ {\Big {[}}\
{\frac {1}{2}}\ m\ {\dot {q}}^{2}(t')\ -\ {\nu}\ m\ (1\ -\ c)\ {\dot
{q}}(t')\ -\ V[q(t'),\ t']\ -\ {\frac  {{\hbar}^{2}}{4\ m\ a^{2}(t')}}\
{\Big {]}}\ +$}
\end{center}
\begin{center}
{$+\ {\frac {m\ v_{o}\ x_{o}}{{\hbar}}}\ +\ {\frac {m\ {\dot
{q}}(t)}{{\hbar}}}\ [x\ -\ q(t)]\ +\ {\frac {m}{2\ {\hbar}}}\ {\Big
{[}}\ {\frac {{\dot {a}}(t)}{a(t)}}\ -\ {\nu}\ c\ {\Big {]}}\
[x\ -\ q(t)]^{2}$\ .\ \ \ \ \ (3.20)}
\end{center}
\par
The above result permit us, finally, to obtain the wave packet
for the linearized S\"{u}ssmann-Hasse-Albrecht-Kostin-Nassar equation
along a classical trajetory. Indeed, considering the eqs. (2.3, 8) and
(3.1, 20), we get:$^{9}$
\begin{center}
{${\psi}(x,\ t)\ =\ [2\ {\pi}\ a^{2}(t)]^{-\ 1/4}\ exp\ {\Big {[}}\ {\Big
{(}}\ {\frac {i\ m}{2\ {\hbar}}}\ [{\frac {{\dot
{a}}(t)}{a(t)}}\ -\ {\nu}\ c]\ -\ {\frac {1}{4\ a^{2}(t)}}\
{\Big {)}}\ [x\ -\ q(t)]^{2}\ {\Big {]}}\ {\times}$}
\end{center}
\begin{center}
{${\times}\ exp\ {\Big {[}}\ {\frac {i\ m\ {\dot {q}}(t)}{{\hbar}}}\ [x\
-\ q(t)]\ +\ {\frac {i\ m\ v_{o}\ x_{o}}{{\hbar}}}\ {\Big {]}}\ {\times}$}
\end{center}
\begin{center}
{${\times}\ exp\ {\Big {[}}\ {\frac {i}{{\hbar}}}\ {\int}_{o}^{t}\ dt'\
{\Big {(}}\ {\frac {1}{2}}\ m\ {\dot {q}}^{2}(t')\ -\ m\ {\nu}\ (1\ -\
c)\ {\dot {q}}(t)\ -\ V[q(t'),\ t']\ -\ {\frac {{\hbar}^{2}}{4\ m\
a^{2}(t')}}\ {\Big {)}}\ {\Big {]}}$\ .\ \ \ \ \ (3.21)}
\end{center}
\par
Note that putting ${\nu}\ =\ 0$ into (3.21) we obtain the
quantum wave packet of the Schr\"{o}dinger equation with the potential
V(x,\ t).$^{[1]}$
\par
4.\ {\bf The Feynman-de Broglie-Bohm Propagator of the Linearized
Schuch-Chung-Hartmann Equation along a Classical Trajetory}
\par
\vspace{0.2cm}
\par
4.1.\ {\bf Introduction}
\vspace{0.2cm}
\par
In 1948,$^{[12]}$ Feynman formulated the following principle of minimum
action for the Quantum Mechanics:
\begin{center}
{{\it The transition amplitude between the states ${\mid}\ a\ >$ and
${\mid}\ b\ >$ of a quantum-mechanical system is given by the sum of
the elementary contributions, one for each trajectory passing by
${\mid}\ a\ >$ at the time t$_{a}$ and by ${\mid}\ b\ >$ at the time
t$_{b}$. Each one of these contributions have the same modulus, but its
phase is the classical action S$_{c{\ell}}$ for each trajectory.}}
\end{center}
\par
This principle is represented by the following expression known as the
"Feynman propagator":
\begin{center}
{$K(b,\ a)\ =\ {\int}_{a}^{b}\ e^{{\frac {i}{{\hbar}}}\ S_{c{\ell}}(b,\
a)}\ D\ x(t)$\ ,\ \ \ \ \ (4.1)}
\end{center}
with:
\begin{center}
{$S_{c{\ell}}(b,\ a)\ =\ {\int}_{t_{a}}^{t_{b}}\ L\ (x,\ {\dot {x}},\
t)\ dt$\ ,\ \ \ \ \ (4.2)}
\end{center}
where $L(x,\ {\dot {x}},\ t)$ is the Lagrangean and $D\ x(t)$ is the
Feynman's Measurement. It indicates that we must perform the integration
taking into account all the ways connecting the states ${\mid}\ a\ >$
and ${\mid}\ b\ >$.
\par
Note that the integral which defines $K(b,\ a)$\ is called "path
integral" or "Feynman integral" and that the Schr\"{o}dinger
wavefunction ${\psi}(x,\ t)$ of any physical system is given by (we
indicate the initial position and initial time by $x_{o}$ and $t_{o}$,
respectively):$^{[13]}$
\begin{center}
{${\psi}(x,\ t)\ =\ {\int}_{-\ {\infty}}^{+\ {\infty}}\ K(x,\ x_{o},\
t,\ t_{o})\ {\psi}(x_{o},\ t_{o})\ dx_{o}$\ ,\ \ \ \ \ (4.3)}
\end{center}
with the quantum causality condition:
\begin{center}
{${\lim\limits_{t,\ t_{o}\ {\to}\ 0}}\ K(x,\ x_{o},\ t,\ t_{o})\ =\
{\delta}(x\ -\ x_{o})$\ .\ \ \ \ \ (4.4)}
\end{center}
\vspace{0.2cm}
\par
4.2.\ {\bf Calculation of the Feynman-de Broglie-Bohm Propagator
for the Li\-nearized S\"{u}ssmann-Hasse-Albrecht-Kostin-Nassar equation
along a Classical Trajetory}
\vspace{0.2cm}
\par
According to Section 3, the wavefunction ${\psi}(x,\ t)$ that was
named wave packet of the of the linearized
S\"{u}ssmann-Hasse-Albrecht-Kostin-Nassar equation along a classical
trajetory, can be written as [see (3.21)]:
\begin{center}
{${\psi}(x,\ t)\ =\ [2\ {\pi}\ a^{2}(t)]^{-\ 1/4}\ exp\ {\Big {[}}\ {\Big
{(}}\ {\frac {i\ m}{2\ {\hbar}}}\ [{\frac {{\dot
{a}}(t)}{a(t)}}\ -\ {\nu}\ c]\ -\ {\frac {1}{4\ a^{2}(t)}}\
{\Big {)}}\ [x\ -\ q(t)]^{2}\ {\Big {]}}\ {\times}$}
\end{center}
\begin{center}
{${\times}\ exp\ {\Big {[}}\ {\frac {i\ m\ {\dot {q}}(t)}{{\hbar}}}\ [x\
-\ q(t)]\ +\ {\frac {i\ m\ v_{o}\ x_{o}}{{\hbar}}}\ {\Big {]}}\ {\times}$}
\end{center}
\begin{center}
{${\times}\ exp\ {\Big {[}}\ {\frac {i}{{\hbar}}}\ {\int}_{o}^{t}\ dt'\
{\Big {(}}\ {\frac {1}{2}}\ m\ {\dot {q}}^{2}(t')\ -\ m\ {\nu}\ (1\ -\
c)\ {\dot {q'}}(t)\ -\ V[q(t'),\ t']\ -\ {\frac {{\hbar}^{2}}{4\ m\
a^{2}(t')}}\ {\Big {)}}\ {\Big {]}}$\ .\ \ \ \ \ (4.5)}
\end{center}
where [see (3.15,16)]:
\begin{center}
{${\ddot {q}}(t)\ +\ {\nu}\ c\ {\dot {q}}(t)\ +\ {\frac {1}{m}}\ V'[q(t),\
t]\ =\ 0$,\ \ \ \ \ (4.6)}
\end{center}
\begin{center}
{${\ddot {a}}(t)\ +\ a(t)\ {\Big {(}}\ {\frac {1}{m}}\ V"[q(t),\ t]\
\ -\ {\nu}^{2}\ c^{2}\ {\Big {)}}\ =\ {\frac {{\hbar}^{2}}{4\
m^{2}\ a^{3}(t)}}$\ .\ \ \ \ \ (4.7)}
\end{center}
where the following initial conditions were obeyed [see (3.17a-d)]:
\begin{center}
{$q(0)\ =\ x_{o}\ ,\ \ \ {\dot {q}}(0)\ =\ v_{o}\ ,\ \ \ a(0)\
=\ a_{o}\ ,\ \ \ {\dot {a}}(0)\ =\ b_{o}$\ .\ \ \ \ \ \
(4.8a-d)}
\end{center}
\par
Therefore, considering (4.3), the Feynman-de Broglie-Bohm propagator
will be calculated using (4.5), in which we will put with no loss of
generality, $t_{o}\ =\ 0$. Thus:
\begin{center}
{${\psi}(x,\ t)\ =\ {\int}_{-\ {\infty}}^{+\ {\infty}}\ K(x,\ x_{o},\
t)\ {\psi}(x_{o},\ 0)\ dx_{o}$\ .\ \ \ \ \ (4.9)}
\end{center}
\par
Let us initially define the normalized quantity:
\begin{center}
{${\Phi}(v_{o},\ x,\ t)\ =\ (2\ {\pi}\ a_{o}^{2})^{1/4}\ {\psi}(v_{o},\
x,\ t)$\ ,\ \ \ \ \ (4.10)}
\end{center}
which satisfies the following completeness relation:$^{[13]}$
\begin{center}
{${\int}_{-\ {\infty}}^{+\ {\infty}}\ dv_{o}\ {\Phi}^{*}(v_{o},\ x,\
t)\ {\Phi}(v_{o},\ x',\ t)\ =\ ({\frac {2\ {\pi}\ {\hbar}}{m}})\
{\delta}(x\ -\ x')$\ .\ \ \ \ \ (4.11)}
\end{center}
\par
Taking the eqs. (2.3,8), we have:
\begin{center}
{${\psi}^{*}(x,\ t)\ {\psi}(x,\ t)\ =\ {\phi}^{2}\ =\ {\rho}(x,\ t)$\
.\ \ \ \ \ (4.12)}
\end{center}
\par
Now, using the eqs. (4.10,12), we get:
\begin{center}
{${\Phi}^{*}(v_{o},\ x,\ t)\ {\psi}(v_{o},\ x,\ t)\ =$}
\end{center}
\begin{center}
{$=\ (2\ {\pi}\ a_{o}^{2})^{1/4}\ {\psi}^{*}(v_{o},\ x,\ t)\
{\psi}(v_{o},\ x,\ t)\ =\ (2\ {\pi}\ a_{o}^{2})^{1/4}\ {\rho}(v_{o},\
x,\ t)\ \ \ {\to}$}
\end{center}
\begin{center}
{${\rho}(v_{o},\ x,\ t)\ =\ (2\ {\pi}\ a_{o}^{2})^{-\ 1/4}\
{\Phi}^{*}(v_{o},\ x,\ t)\ {\psi}(v_{o},\ x,\ t)$\ .\ \ \ \ \
(4.13)}
\end{center}
\par
On the other side, substituting (4.13) into (2.20), integrating the
result and using (3.1) and (4.10) results [remembering
that ${\psi}^{*}\ {\psi}({\pm}\ {\infty})\ \ \ {\to}\ \ \ 0$]:
\begin{center}
{${\frac {{\partial}({\Phi}^{*}\ {\psi})}{{\partial}t}}\ +\ {\frac
{{\partial}({\Phi}^{*}\ {\psi}\ {\vartheta}_{qu})}{{\partial}x}}\ =\ 0\
\ \ {\to}$}
\end{center}
\begin{center}
{${\frac {{\partial}}{{\partial}t}}\ {\int}_{-\ {\infty}}^{+\ {\infty}}\
dx\ {\Phi}^{*}\ {\psi}\ +\ ({\Phi}^{*}\ {\psi}\
{\vartheta}_{qu}){\mid}_{-\ {\infty}}^{+\ {\infty}}\ =$}
\end{center}
\begin{center}
{$=\ {\frac {{\partial}}{{\partial}t}}\ {\int}_{-\ {\infty}}^{+\ {\infty}}\
dx\ {\Phi}^{*}\ {\psi}\ +\ (2\ {\pi}\ a_{o}^{2})^{1/4}\ ({\psi}^{*}\
{\psi}\ {\vartheta}_{qu}){\mid}_{-\ {\infty}}^{+\ {\infty}}\ =\ 0\ \ \
{\to}$}
\end{center}
\begin{center}
{${\frac {{\partial}}{{\partial}t}}\ {\int}_{-\ {\infty}}^{+\
{\infty}}\ dx\ {\Phi}^{*}\ {\psi}\ =\ 0$\ .\ \ \ \ \ (4.14)}
\end{center}
\par
The eq. (4.14) shows that the integration is time independent.
Consequently:
\begin{center}
{${\int}_{-\ {\infty}}^{+\ {\infty}}\ dx'\ {\Phi}^{*}(v_{o},\ x',\ t)\
{\psi}(x',\ t)\ =\ {\int}_{-\ {\infty}}^{+\ {\infty}}\ dx_{o}\
{\Phi}^{*}(v_{o},\ x_{o},\ 0)\ {\psi}(x_{o},\ 0)$\ .\ \ \ \ \ (4.15)}
\end{center}
\par
Multiplying (4.15) by ${\Phi}(v_{o},\ x,\ t)$ and integrating over
$v_{o}$ and using (4.11), we obtain [remembering
that ${\int}_{-\ {\infty}}^{+\ {\infty}}\ dx'\ f(x')\ {\delta}(x' -\
x)\ = f(x)$]:
\begin{center}
{${\int}_{-\ {\infty}}^{+\ {\infty}}\ {\int}_{-\ {\infty}}^{+\
{\infty}}\ dv_{o}\ dx'\ {\Phi}(v_{o},\ x,\ t)\ {\Phi}^{*}(v_{o},\ x',\ t)\
{\psi}(x',\ t)$\ =}
\end{center}
\begin{center}
{=\ ${\int}_{-\ {\infty}}^{+\ {\infty}}\ {\int}_{-\
{\infty}}^{+\ {\infty}}\ dv_{o}\ dx_{o}\ {\Phi}(v_{o},\ x,\ t)\
{\Phi}^{*}(v_{o},\ x_{o},\ 0)\ {\psi}(x_{o},\ 0)\ \ \ {\to}$}
\end{center}
\begin{center}
{${\int}_{-\ {\infty}}^{+\ {\infty}}\ dx'\ ({\frac {2\ {\pi}\
{\hbar}}{m}})\ {\delta}(x'\ -\ x)\ {\psi}(x',\ t)\ =\ ({\frac {2\ {\pi}\
{\hbar}}{m}})\ {\psi}(x,\ t)$\ =}
\end{center}
\begin{center}
{=\ ${\int}_{-\ {\infty}}^{+\ {\infty}}\ {\int}_{-\ {\infty}}^{+\
{\infty}}\ dv_{o}\ dx_{o}\ {\Phi}(v_{o},\ x,\ t)\ {\Phi}^{*}(v_{o},\
x_{o},\ 0)\ {\psi}(x_{o},\ 0)\ \ \ {\to}$}
\end{center}
\begin{center}
{${\psi}(x,\ t)\ =\ {\int}_{-\ {\infty}}^{+\ {\infty}}\ {\Big {[}}\
({\frac {m}{2\ {\pi}\ {\hbar}}})\ {\int}_{-\ {\infty}}^{+\ {\infty}}\
dv_{o}\ {\Phi}(v_{o},\ x,\ t)\ {\times}$}
\end{center}
\begin{center}
{${\times}\ {\Phi}^{*}(v_{o},\ x_{o},\ 0)\ {\Big {]}}\ {\psi}(x_{o},\
0)\ dx_{o}$\ .\ \ \ \ \ (4.16)}
\end{center}
\par
Comparing (4.9) and (4.16), we have:
\begin{center}
{$K(x,\ x_{o},\ t)\ =\ {\frac {m}{2\ {\pi}\ {\hbar}}}\ {\int}_{-\
{\infty}}^{+\ {\infty}}\ dv_{o}\ {\Phi}(v_{o},\ x,\ t)\
{\Phi}^{*}(v_{o},\ x_{o},\ 0)$\ .\ \ \ \ \ (4.17)}
\end{center}
\par
Substituting (4.5) and (4.10) into (4.17), we finally obtain the
Feynman-de Broglie-Bohm Propagator of the linearized
S\"(u)ssmann-Hasse-Albrecht-Kostin-Nassar equation along a classical
trajetory [remembering that ${\Phi}^{*}(v_{o},\ x_{o},\ 0)\ =\ exp\ (-\
{\frac {i\ m\ v_{o}\ x_{o}}{{\hbar}}})$]:
\begin{center}
{$K(x,\ x_{o};\ t)\ =\ {\frac {m}{2\ {\pi}\ {\hbar}}}\ {\int}_{-\
{\infty}}^{+\ {\infty}}\ dv_{o}\ {\sqrt {{\frac
{a_{o}}{a(t)}}}}\ {\times}$}
\end{center}
\begin{center}
{${\times}\ exp\ {\Big {[}}\ {\Big
{(}}\ {\frac {i\ m}{2\ {\hbar}}}\ [{\frac {{\dot
{a}}(t)}{a(t)}}\ -\ {\nu}\ c]\ -\ {\frac {1}{4\ a^{2}(t)}}\
{\Big {)}}\ [x\ -\ q(t)]^{2}\ +\ {\frac {i\ m\ {\dot
{q}}(t)}{{\hbar}}}\ [x\ -\ q(t)]\ {\Big {]}}\ {\times}$}
\end{center}
\begin{center}
{${\times}\ exp\ {\Big {[}}\ {\frac {i}{{\hbar}}}\
{\int}_{o}^{t}\ dt'\ {\Big {(}}\ {\frac {1}{2}}\ m\ {\dot
{q}}^{2}(t')\ -\ m\ {\nu}\ (1\ -\ c)\ {\dot {q}}(t')\ -\ V[q(t'),\ t']\
-\ {\frac {{\hbar}^{2}}{4\ m\ a^{2}(t')}}\ {\Big {)}}\ {\Big {]}}$\ ,\
\ \ (4.18)}
\end{center}
where $q(t)$ and $a(t)$ are solutions of the (4.6,\ 7) differential
equations.
\par
Finally, it is important to note that putting ${\nu}\ =\ 0$ and
V[q(t'),\ t']\ =\ 0 into (4.6), (4.7) and (4.18) we obtain the free
particle Feynman propagator.$^{[1,\ 14]}$
\par
\begin{center}
{{\bf NOTES AND REFERENCES}}
\end{center}
\par
1.\ BASSALO, J. M. F., ALENCAR, P. T. S., CATTANI, M. S. D. e
NASSAR, A. B. {\it T\'opicos da Mec\^anica Qu\^antica de de
Broglie-Bohm}, {\bf E-Book}\ ({\it
http://www-sbi.if.usp.br/?q=note/117}-1655) (2010).
\par
2.\ S\"{U}SSMANN, D. {\it Seminar Talhk at Los Alamos} (1973).
\par
3.\ HASSE, R. W. {\it Journal of Mathematical Physics 16}, 2005
(1975).
\par
4.\ ALBRECHT, K. {\it Physics Letters B56}, 127 (1975).
\par
5.\ KOSTIN, M. D. {\it Journal of Statistical Physics 12}, 146 (1975).
\par
6.\ NASSAR, A. B. {\it Journal of Mathematical Physics 27}, 2949 (1986).
\par
7.\ MADELUNG, E. {\it Zeitschrift f\"{u}r Physik 40}, 322 (1926).
\par
8.\ BOHM, D. {\it Physical Review 85}, 166 (1952).
\par
9.\ BASSALO, J. M. F., ALENCAR, P. T. S., SILVA, D. G., NASSAR, A. B.
and CATTANI, M. {\it arXiv:0905.4280v1}\ [quant-ph]\ 26\ May\
2009; -----. {\it arXiv:1004.1416v1}\ [quant-ph]\ 10\ April\
2010; -----. {\it arXiv:1006.1868v1}\ [quant-ph]\ 9\ June\
2010; -----. {\it arXiv:1010.2640v1}\ [quant-ph]\ 13\ October\
2010.
\par
10.\ NASSAR, A. B., BASSALO, J. M. F., ALENCAR, P. T. S., CANCELA, L. S.
G. and CATTANI, M. {\it Physical Review 56E}, 1230 (1997).
\par
11.\ NASSAR, A. B. {\it Journal of Mathematics Physics 27}, 2949
(1986).   
\par
12.\ FEYNMAN, R. P. {\it Reviews of Modern Physics 20}, 367 (1948).
\par
13.\ BERNSTEIN, I. B. {\it Physical Review A32}, 1 (1985).
\par
14.\ FEYNMAN, R. P. and HIBBS, A. R. {\it Quantum Mechanics and Path
Integrals}, McGraw-Hill Book Company (1965).
\end{document}